\documentclass[aps,prb,twocolumn,groupedaddress]{revtex4}

\usepackage{amsfonts}
\usepackage{graphicx}
\begin{document}

\title{Experimental realization of a relativistic fluxon ratchet}

\author{G. Carapella}
\thanks{Corresponding author}
\email[\newline e-mail: ]{giocar@sa.infn.it}
\thanks{\newline FAX: +3908965390 \\}

\affiliation{Departement of Physics \lq\lq E. R. Caianiello\rq\rq
, and INFM Research Unit, University of Salerno, via
S. Allende, I-84081 Baronissi, Italy. }
\author{G. Costabile}
\affiliation{Departement of Physics \lq\lq E. R. Caianiello\rq\rq
, and INFM Research Unit, University of Salerno, via
S. Allende, I-84081 Baronissi, Italy. }

\author{R. Latempa}
\affiliation{Departement of Physics \lq\lq E. R. Caianiello\rq\rq
, and INFM Research Unit, University of Salerno, via
S. Allende, I-84081 Baronissi, Italy. }

\author{N. Martucciello}
\affiliation{Departement of Physics \lq\lq E. R. Caianiello\rq\rq
, and INFM Research Unit, University of Salerno, via
S. Allende, I-84081 Baronissi, Italy. }

\author{M. Cirillo}
\affiliation{
INFM and Department of Physics,
        University of Roma "Tor Vergata", I-00173 Rome, Italy. }

\author{A. Polcari}
\affiliation{INSTM  and Department of Physics,
        University of Salerno, I-84081 Baronissi, Italy}

\author{G. Filatrella}
\affiliation{University of Sannio and INFM Research Unit, University of Salerno, via
S. Allende, I-84081 Baronissi, Italy. }

\date{\today}

\begin{abstract}
We report the observation of the ratchet effect for a
relativistic flux quantum trapped in an annular Josephson junction embedded
in an inhomogeneous magnetic field.
In such a solid state system mechanical
quantities are proportional to electrical quantities, so that the
ratchet effect
represents the realization of a relativistic-flux-quantum-based diode. Mean
static voltage response, equivalent to directed fluxon motion,
is experimentally
demonstrated in such a  diode
for deterministic current forcing both in the overdamped and in the
underdamped dynamical regime.  In the underdamped regime, the
recently predicted phenomenon of current reversal is also recovered in our
fluxon ratchet.

\end{abstract}


\maketitle

%\begin{multicols}{2}
%\narrowtext


\section{Introduction}
A particle in a periodic potential lacking spatial reflection symmetry,
known as a ratchet potential \cite{Mag}, exhibits a net unidirectional
motion in the absence of a net driving force. This static response
to an oscillating
force is known as the ratchet effect.
 The net
unidirectional motion exhibited in ratchet potentials is the key feature
potentially interesting for applications.  In Josephson
junction systems, various realizations of ratchet
effect have been investigated \cite{Zap,Wei,Tri,Gol}.

Recently we
proposed \cite{GC} a single flux quantum in a
long annular Josephson
junction embedded in an inhomogeneous magnetic field as
an experimentally
accessible
solid state example of a  relativistic particle in a
ratchet potential. 
Here we briefly review  the experimental
demonstration of the ratchet
effect in such a solid state system in the
overdamped regime \cite{GC1} and we
report  new data in the underdamped regime.

\section{Theory}

\begin{figure}[b]
\includegraphics[width=7cm,clip=]{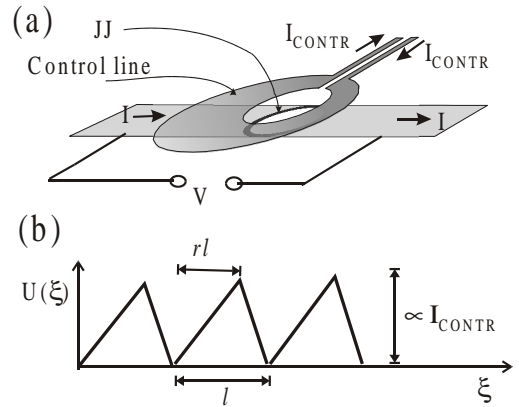} 
\caption{ (a) A long annular Josephson junction with a control line
generating a sawtooth-like magnetic field.
(b) The effective potential
experienced by a flux quantum trapped in the junction
when the control current is turned on.}
\label{fig1}
\end{figure}

An inhomogeneous magnetic field  generates a potential  
for a flux
quantum trapped in  a long annular Josephson junction
[see Fig.~\ref{fig1}(a)].
With a suitable choice \cite{GC}
for the width $W(x)$ of a control line on the top of the junction
[see
Fig.~\ref{fig1}(a)] a
sawtooth-like magnetic field
can be generated.

Using the perturbative approach,
the equation of motion for the center of
mass $\xi (t)$ of a flux quantum trapped
in the junction can be found.
The result \cite{GC} is the equation of a
 relativistic particle subjected to an inhomogeneous force $F(\xi
)$ with a sawtooth-like ratchet potential $U(\xi )$
[see Fig.~\ref{fig1}(b)] and an
homogeneous forcing $\eta $:
\begin{equation}
\frac{4}{\pi }\left( 1-\stackrel{\cdot}{\xi }^{2}\right)
^{-\frac{3}{2}}\stackrel{\cdot \cdot}{\xi }+
\frac{4\alpha }{\pi }\frac{\stackrel{\cdot}{\xi }}{\sqrt{1-\stackrel{\cdot 
}{\xi }^{2}}}
=\eta +F(\xi ).
\label{inert}
\end{equation}
Here the strength
of the ratchet potential $U(\xi )$ is controlled by the
control current $I_{CONTR}$ and the homogeneous forcing $\eta $ by the
current $I$ feeding
the junction. The measured static voltage $V$ is proportional to the spatial
mean $u=<\stackrel{\cdot }{\xi }>_{s}$ of the fluxon velocity. In physical
units, $V=\Phi _{0}u/L$ with limiting velocity $u=\overline{c}.$

\section{Experiments}
The samples were fabricated with the geometry shown
in Fig.~\ref{fig1}(a). The
Nb/Al$_{2}$O$_{3}$/Nb annular junctions
had a circumference $L=$ 1200 $\mu $%
m and width $W=$20 $\mu $m. The control lines were electrically insulated
from the junctions by a 300~nm thick SiO$_{2}$ layer.
Here we report
  junctions with normalized lengths
$l=L/ \lambda_{J}\simeq 10$, and
$l\simeq 20.$

\subsection{Overdamped regime}
\begin{figure}[t]
\includegraphics[width=7cm,clip=]{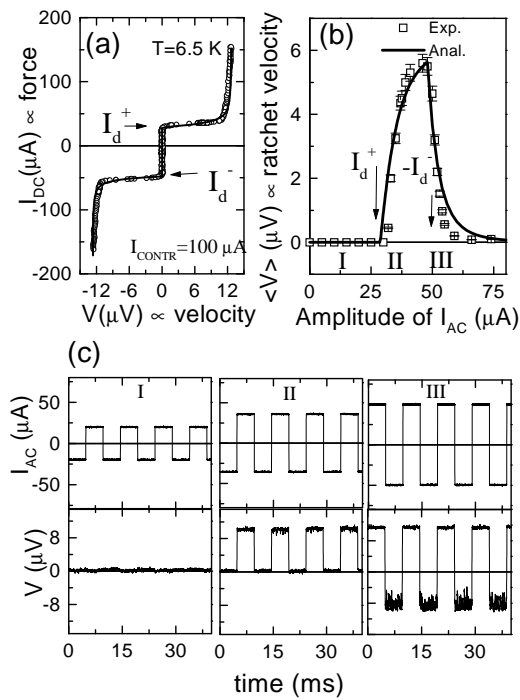}
\caption{(a) Current-voltage  curve
of the single fluxon trapped in the junction. (b) Measured
 ratchet voltage 
induced by a square wave ac forcing in the adiabatic limit.
 (c) Digital oscilloscope time
traces of forcing current and measured voltage for amplitude of forcing
signal in the three regions shown in (b).}
\label{fig2}
\end{figure}

Figure~\ref{fig2}(a) shows the measured
current-voltage (force-velocity) curve
\begin{figure}[t]
\includegraphics[width=7cm,clip=]{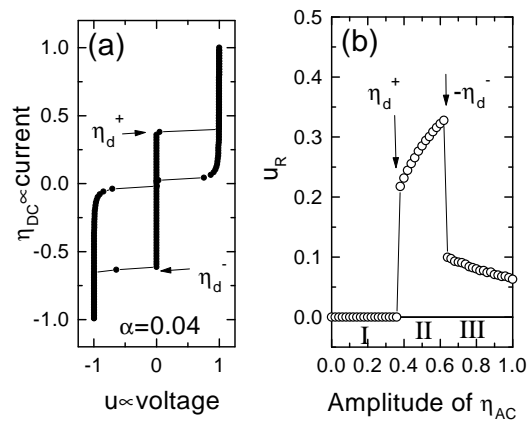}
\caption{ Calculated current-voltage curve (a) and ratchet
voltage (b)
of the fluxon in the underdamped (inertial) limit. In (b) 
a triangular adiabatic ac forcing was used.}
\label{fig3}
\end{figure}

of the fluxon trapped in the junction with
 $l\simeq 10$
 when a
current is fed in the control line. As expected for
an asymmetric potential,
two different depinning current $I_{d}^{+}$ and $I_{d}^{-}$
are recovered.  The curve is recorded at T=6.5~K. At this
temperature, the dissipation parameter $\alpha $ is large enough to
overcome inertial effects and the overdamped regime
with a nonhysteretic current-voltage curve is achieved.
In this regime the dynamics is described by the
noninertial version \cite{GC} of Eq.~(\ref{inert}).

The ratchet effect
generates a net transport velocity, $u_{R},$ in response
to an ac forcing. In our system such a velocity is proportional to the
measured voltage $V_{R}=<V>$ averaged over the period of an
alternating current fed in to 
the junction.  For ac forcing we use an "adiabatic" square wave, i.e.,
 with a period
much larger than the typical response time of the system,  of the order
of 150~ps.  The measured ratchet
voltage as a function of the amplitude of the ac forcing is
plotted in Fig.~\ref{fig2}(b).
A comparison of experimental data \cite{GC1} with theory \cite{GC} is
also shown.

The three typical regions expected from the ratchet
effect are clearly exhibited: the static region I where the fluxon is at rest
(pinned), the active region II where the fluxon motion is strictly
unidirectional, and the overdriven region III where fluxon motion is
unidirectional in the mean \cite{GC}. 
From the
time
domain snapshots in 
Fig.~\ref{fig3}(c),
we also notice a strong
rectification of the input current (diode effect) in the active region 
and a weak rectification effect in the overdriven region.

\subsection{Underdamped regime and current reversal effect}

\begin{figure}[t]
\includegraphics[width=7cm,clip=]{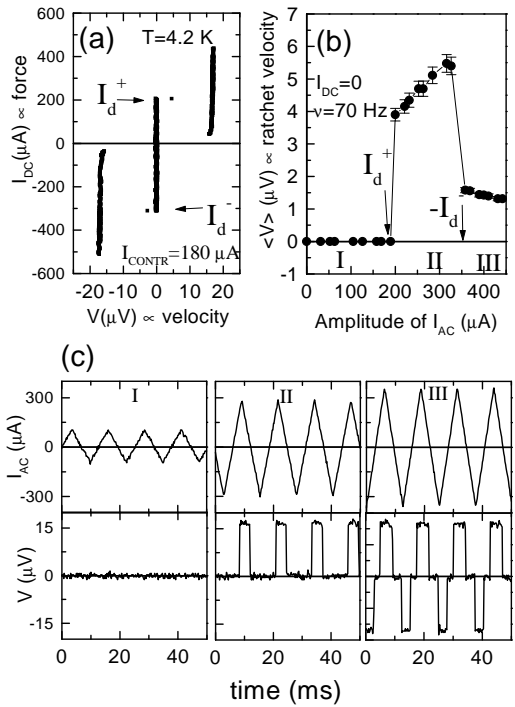}
\caption{ Same meaning as Fig.~\ref{fig2}. Here the
temperature of the thermal bath is lowered to achieve
the underdamped (inertial) limit and a triangular wave was
chosen as ac adiabatic forcing.}
\label{fig4}
\end{figure}

As noted above, results in Fig.~{\ref{fig2}} were
recorded in the overdamped regime we achieved at T=6.5~K.
As the thermal bath temperature  is lowered,
 the dissipation parameter $\alpha $ is reduced more and more, so that an
underdamped regime with
dynamics described by Eq.~(\ref{inert}) can be achieved.
The numerically calculated current-voltage curve 
for this regime, shown in Fig.~\ref{fig3}(a)
is now fully hysteretic. The ratchet voltage calculated for
an adiabatic triangular ac current forcing is shown in
Fig.~\ref{fig3}(b).
Experimentally, such an inertial regime is achieved at
T=4.2~K for our fluxon. Both the measured current-voltage
[Fig.~\ref{fig4}(a)] and the
measured ratchet voltage [Fig.~\ref{fig4}(b)] fully agree with
numerical predictions shown in Fig.~\ref{fig3}. 
The
time
domain snapshots in 
Fig.~\ref{fig4}(c)
suggest that  the
 strongest
rectification of the input current (diode effect) is again
achieved in
the active region. Moreover,
with respect to the noninertial regime,
a substantial rectification is also obtained in the
overdriven region.

As stated above, our fluxon has a typical response time of
the order of 150~ps, so that a response to microwave signals
can be addressed.
Theory  \cite{GC}
predicts  that  when a.c. current forcing with frequency
$\nu _{RF}$ in the
microwave range is used, the d.c. characteristic ($I_{DC}-V$) should
exhibits current steps at voltages $V_{n}=n\Phi _{0}\nu _{RF}$, with
$n$ integer and $\Phi _{0}$ the flux-quantum.
Numerically, the voltage spacing of these microwave induced
steps is
\begin{equation}
\Delta V[\mu {\rm V}]=2.07 \nu _{RF}[{\rm GHz}].
\label{deltav}
\end{equation}
Dynamically, these curent steps accounts for a locking
of the revolution motion of the fluxon with the external drive \cite{GC}. 
Moreover, the ratchet effect, i.e., the existence of a mean
voltage as a response to an a.c. forcing, compels a zero-current-axis
crossing \cite{GC} in the $I_{DC}-V$ curve. In other words,
one should observe $V\neq $ 0 at $I_{DC}=0$ when the junction
is irradiated with microwaves.

The mean voltage  $V(I_{DC}=0)\equiv V_{R}$ induced by the a.c. current
forcing normally exibits only one polarity for a fixed sign
of the static spatial ratchet potential.
For example in  Fig.~\ref{fig2}(b), and Fig~\ref{fig4}(b) a positive
polarity was recovered for the used potential.
This means that the mean velocity of the fluxon was
always positive (or zero, in the static region).
However, recently has been predicted
\cite{Mateos,Barbi} that the ratchet velocity can
change sign in the inertial regime.
This fenomenon is  known as
\begin{figure}[t]
\includegraphics[width=7cm,clip=]{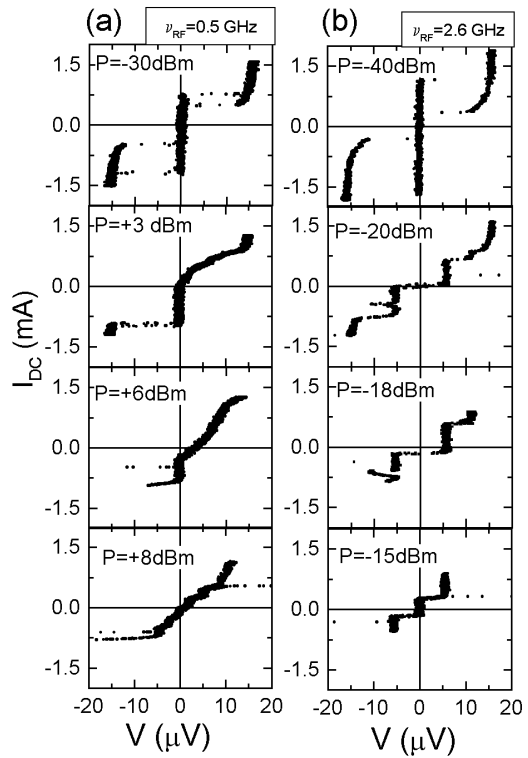}
\caption{
 Modification of the $I_{DC}-V$ curve of the induced by
a microwave signal at frequency
$\nu _{RF}$=0.5~GHz (a) and
$\nu _{RF}$=2.6~GHz (b). A zero current axis crossing
($V\neq $ 0 at $I_{DC}=0$)
is
observed, togheter with locked current steps.}
\label{fig5}
\end{figure}

"current reversal" effect. The word "current" is mutuated
from brownian motion field and means net velocity.
In our system this correspond to our ratchet voltage $V_{R}$.
Current reversal (inversion of $V_{R}$ in our system) can appear
for frequencies of the a.c. forcing comparable with
the typical frequencies of evolution of the inertial ratchet system,
or, in other words, for frequencies where locking occurs.
In our system this means the microwave range.
Dynamically, the current reversal corresponds to the onset of a
 chaotic dynamical regime \cite{Mateos,Barbi}.

\begin{figure}[t]
\includegraphics[width=7cm,clip=]{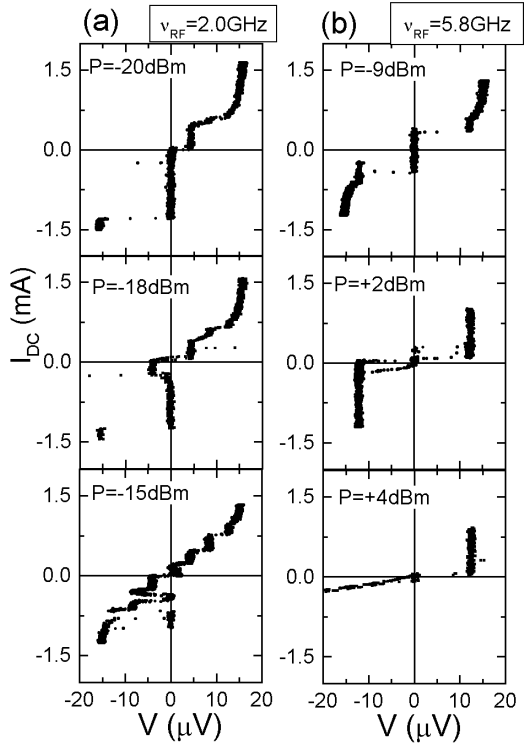}
\caption{ $I_{DC}-V$ curves for
microwave signal at frequency
$\nu _{RF}$=2.0~GHz (a) and
$\nu _{RF}$=5.8~GHz (b) showing the
current reversal phenomenon
($V< $ 0 at $I_{DC}=0$).}
\label{fig6}
\end{figure}

Experiments with microwave forcing currents were performed
on the junction with $l\simeq 20$. The control current
was chosen to have a (normally) positive ratchet voltage.
The modification of the $I_{DC}-V$ curve for increasing
power of a signal at $\nu _{RF}$=0.5~GHz is shown in Fig.~\ref{fig5}(a).
As expected \cite{GC}, a zero current
axis crossing is observed ($V_{R} > 0$ at P=~6dBm) with manifestation
of small current steps
 spaced about 1~$\mu $V, as expected from relation (\ref{deltav}).
The current steps are better evident
in Fig.~\ref{fig5}(b), where a $\nu _{RF}$=2.6~GHz  was used.
Consistently with (\ref{deltav}) here $\Delta V \approx 5.4~ \mu {\rm V}$
is observed.
Moreover, also somewhat that recall
a chaotic behavior is envisaged in the current-voltage curves (at P=-20~dBm
and P=-18~dBm),
but no current reversal is observed at the chosen frequency
(is always $V_{R} \ge 0$).
Such a current reversal ($V_{R} <  0$) is instead
observed with  $\nu _{RF}$=2.0~GHz and  $\nu _{RF}$=5.8~GHz, as
shown in  Fig.~\ref{fig6}.
The current steps corresponding to a motion syncronized
in the direction opposite to the
"natural" direction, i.e., the steps with negative voltage
polarity in Fig.~\ref{fig6}, are quite noisy and instable,
accounting for some quite chaotic internal dynamics.
The onset of a chaotic dynamics is also envisaged
when current reversal occurs, as better seen
in the panels at P=-18 dBm in Fig.~\ref{fig6}(a), and
at P=2~dBm in Fig.~\ref{fig6}(b). This is in qualitative agreement
with theory \cite{Mateos,Barbi} of inertial ratchets.

\section{Conclusions}

We have considered a single flux quantum in a
long annular Josephson
junction embedded in an inhomogeneous magnetic field as
an experimentally
accessible
solid state example of a  relativistic particle in a
ratchet potential.
In such a solid state system mechanical
quantities are proportional to electrical quantities, so that the
ratchet effect
represents the realization of a relativistic-flux-quantum-based diode.
 Mean
static voltage response, equivalent to directed fluxon motion,
has been experimentally
demonstrated in such a  diode
for deterministic and adiabatic current forcing
both in the overdamped (noninertial)
and in the
underdamped (inertial) dynamical regime.  In the underdamped regime,
also nonadiabatic current forcing, corresponding to the
microwave range forcing, has been experimentally addressed, and
the predicted synchronization with external drive as well as the
 current reversal phenomenon have been
 recovered in the system.

\section*{Acknowledgements}
This work was partially supported by MURST COFIN00
project.

\end{document}